\documentclass[aps,pre,twocolumn,showpacs,superscriptaddress]{revtex4}
\usepackage{epsfig}
\usepackage{times}
\begin{document}

\title{Impact of generalized benefit functions on the evolution of cooperation in spatial public goods games with continuous strategies}

\author{Xiaojie Chen}
\email{chenx@iiasa.ac.at}
\affiliation{Evolution and Ecology Program, International Institute for Applied Systems Analysis (IIASA), Schlossplatz 1, A-2361 Laxenburg, Austria}
\author{Attila Szolnoki}
\affiliation{Institute of Technical Physics and Materials Science, Research Centre for Natural Sciences, Hungarian Academy of Sciences, P.O. Box 49, H-1525 Budapest, Hungary}
\author{Matja\v{z} Perc}
\affiliation{Faculty of Natural Sciences and Mathematics, University of Maribor, Koro{\v s}ka cesta 160, SI-2000 Maribor, Slovenia}
\author{Long Wang}
\affiliation{State Key Laboratory for Turbulence and Complex Systems, College of Engineering, Peking University, Beijing, China}

\begin{abstract}
Cooperation and defection may be considered as two extreme responses to a social dilemma. Yet the reality is much less clear-cut. Between the two extremes lies an interval of ambivalent choices, which may be captured theoretically by means of continuous strategies defining the extent of the contributions of each individual player to the common pool. If strategies are chosen from the unit interval, where $0$ corresponds to pure defection and $1$ corresponds to the maximal contribution, the question is what is the characteristic level of individual investments to the common pool that emerges if the evolution is guided by different benefit functions. Here we consider the steepness and the threshold as two parameters defining an array of generalized benefit functions, and we show that in a structured population there exist intermediate values of both at which the collective contributions are maximal. However, as the cost-to-benefit ratio of cooperation increases the characteristic threshold decreases, while the corresponding steepness increases. Our observations remain valid if more complex sigmoid functions are used, thus reenforcing the importance of carefully adjusted benefits for high levels of public cooperation.
\end{abstract}

\maketitle

\section{Introduction}
The public goods game is a typical example of an evolutionary game \cite{hofbauer_98, nowak_06, sigmund_10} that is governed by group interactions. It requires that players decide simultaneously whether they wish to bare the cost of cooperation and thus to contribute to the common pool, or not. Regardless of their decision, each member of the group receives an equal share of the public good after the initial contributions are multiplied by a benefit factor that takes into account the added value of collaborative efforts. Individuals are best off by not contributing anything to the common pool, \textit{i.e.} by defecting, while the group is most successful if everybody invests to the common pool, \textit{i.e.} cooperates. Since the interests of individual players evidently do not agree with the interests of the group as a whole, we have a blueprint of a social dilemma that threatens to evolve towards the ``tragedy of the commons'' \cite{hardin_g_s68}. While the impetus of prosocial behavior in settings described by the public goods game is commonly attributed to between-group conflicts \cite{bowles_11} and alloparental care \cite{hrdy_11}, mechanisms that might facilitate and maintain highly cooperative states are still sought ardently \cite{nowak_11}.

Public goods are particularly vulnerable to exploitation since group interactions that bring them about tend to blur the traces of those that defect. Reciprocity \cite{trivers_qrb71, nowak_n98} for example, \textit{i.e.} the act of returning favor for a favor, is straightforward in games governed by pairwise interactions, but becomes problematic in games governed by group interactions. The same is true for punishment \cite{sigmund_tee07}, as those that ought to be punished may not be easily traced down. Despite of these well known difficulties associated with the promotion of cooperation in the public goods game, complex interaction networks \cite{santos_n08, gomez-gardenes_c11, roca_pone10, buesser_pa11, gomez-gardenes_epl11, wang_z_epl12}, inhomogeneous player activities \cite{guan_pre07}, appropriate partner selection \cite{wu_t_pre09, zhang_hf_epl11}, diversity \cite{yang_hx_pre09, ohdaira_acs11, perc_njp11}, voluntary participation \cite{hauert_jtb02,szabo_prl02}, heterogeneous wealth distributions \cite{wang_j_pre10b}, the introduction of punishment \cite{brandt_prsb03, helbing_ploscb10, sigmund_n10, szolnoki_pre11, sigmund_dga11,szolnoki_pre11b} and reward \cite{szolnoki_epl10, hauert_jtb10}, risk of collective failures \cite{santos_pnas11}, coordinated investments \cite{vukov_jtb11}, as well as both the joker \cite{arenas_jtb11} and the Matthew effect \cite{perc_pre11} were all recently identified as viable means to avoid the tragedy of the commons in structured populations \cite{zhang_j_pa11, wakano_pnas09, dai_ql_njp10, traulsen_pnas09, szolnoki_pre09c, lin_yt_pa11}.

In the present paper, we depart from the traditionally assumed notion of discrete strategies by taking into account the whole continuous range of the strategy space. That is to say, players are no longer either pure cooperators or defectors, but they can choose between all the possible nuances between these two extremes. Indeed, the continuous version of the public goods game \cite{killingback_prslb99, janssen_jtb06} can be considered an additional step towards more realistic conditions, given that especially humans are unlikely to stick with simply one or the other pure strategy. The transition from the two discrete to a continuous strategy set can be achieved most elegantly by introducing a continuous variable from the unit interval defining the fraction of the total cost a given player is willing to bare. While the limits $0$ and $1$ recover the two pure strategies, intermediate values from the unit interval correspond to more or less cooperative players. An obvious but important distinction from the discrete version of the public goods game is that the continuous version allows for the evolution of an intermediate level of investments from players, which makes it particularly apt for the investigation of the impact of different benefit functions.

The most frequent assumption is that the benefit returned by the public goods game scales linearly with the amount contributed by the cooperators, \textit{i.e.} the more that is contributed the more can be shared. There are situations, however, where this assumption obviously fails and a nonlinear function becomes more appropriate. A prominent example is constituted by the so-called threshold public goods game, where the sum of contributions is multiplied by the benefit factor only if the former exceeds a certain threshold \cite{cadsby_pc08, souza_mo_jtb09, wang_j_pre09}. In case all players are equal this simplifies to the critical mass problem \cite{szolnoki_pre10}. Thresholds, being described by step-like benefit functions, can be considered as an extreme case of a general nonlinear benefit function \cite{boza_bmceb10, deng_k_pone11b}, with the other extreme being when the public good depends only slightly (or not at all) on the contributions of the members. The generalized sigmoid function bridges these two extremes and is characterized by two parameters, namely the threshold and the steepness parameter. Here we consider such a sigmoid benefit function and study how both the steepness and the threshold affect the evolution of cooperation in the spatial public goods game with continuous strategies. Before proceeding with a more accurate description of the model and the presentation of the main results, our conclusions can be briefly summarized as follows: There exists an intermediate value of the steepness and threshold in the sigmoid function, which warrant the evolution of the highest collective efforts of players. Upon increasing the cost-to-benefit ratio, however, the steepness increases, whereas the corresponding threshold value decreases. These results are highly robust to variations in the complexity of the sigmoid function and bolster the importance of benefits for the successful evolution of public cooperation.

\section{Model}
\begin{figure}
\begin{center} \includegraphics[width = 8.4cm]{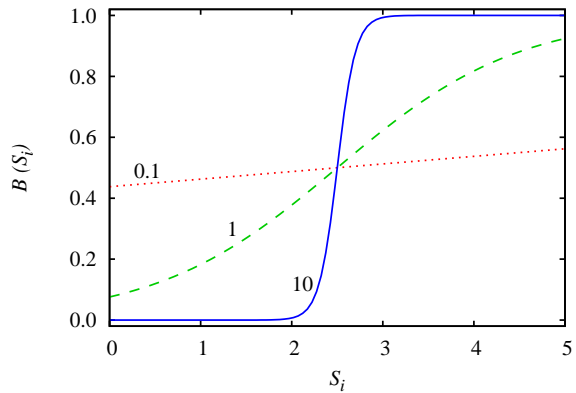}
\caption{\label{benefit}(Color online) Outlays of the benefit function $B(S_i)$ for different values of $\beta$, as indicated on the graph. The threshold value is $T=2.5$.}
\end{center}
\end{figure}

We consider the continuous public goods game on a square lattice of size $L \times L$ with periodic boundary conditions and nearest neighbor interactions. The strategy of each player $x$ is initially drawn uniformly at random from the unit interval $s_x \in [0,1]$, defining its level of contribution in each of the five groups $G_i$ ($i=1,\ldots, 5$) of size $N=5$ where it is member. Accordingly, the total payoff of player $x$ is $P_x=\sum_i P_x^i$, where
\begin{equation}
P_x^i=bB(S_i)-s_x c
\label{payoff}
\end{equation}
is the payoff obtained from group $G_i$. In Eq.~\ref{payoff} $b$ is the benefit of the public good, $c$ $(c<b)$ is the cost of cooperation, $S_i=\sum_{y\in G_i} s_y$ is the total amount of collected contributions while $B(S_i)$ is the benefit function determining the total amount of the produced public good. In order to take into account both extremes, namely when the produced public good depends slightly or heavily on the contributions of group members, the function
\begin{equation}
B(S_i)=\frac{1}{1+\exp[-\beta(S_i-T)]}
\label{nonlinear}
\end{equation}
is used, where $T$ represents the threshold value, and $\beta$ represents the steepness of the function \cite{boza_bmceb10}. For $\beta=0$, the benefit function is a constant equalling $0.5$. In this situation, the public goods are insensitive to the efforts of group members. Conversely, for $\beta=+\infty$ the benefit function becomes step-like so that group members can enjoy the benefits of collaborative efforts via $b$ only if the total amount of contributions in the group $S_i$ exceeds a threshold. Otherwise, they obtain nothing. For clarity, the benefit function $B(S_i)$ is plotted in Fig.~\ref{benefit} for different values of $\beta$.

After playing the game, each player is allowed to learn from one of its neighbors and potentially update its strategy. Player $x$ adopts the strategy $s_y$ from one randomly chosen neighbor $y$ with a probability
\begin{equation}
f=\frac{1}{1+\exp[(P_x-P_y)/\kappa]},
\end{equation}
where $\kappa$ denotes the amplitude of noise \cite{szabo_pr07}. Without losing generality, we set $\kappa=0.5$ so that it is very likely that the better performing players will pass their strategy onto their neighbors, yet it is also possible that players will occasionally learn from those performing worse. We note that the presented results are largely independent on the actual value of noise and remain valid up to $\kappa \approx 3$.

According to the imitation rule player $x$ imitates accurately the strategy of player $y$, which may cause problem during numerical simulations because we have infinitely large number of strategies but only a finite number of players. As a result the final output might depend on the initial condition especially at small system sizes. This problem can be elegantly alleviated if we introduce imitation errors resulting in a slightly different $s_y$ for player $x$. More precisely, the new strategy of player $x$ is $s_x\prime=s_y \pm w \sigma \mid s_x-s_y \mid$, where $\sigma \in [0,1]$ is a random number and $w=0.1$ is a weight factor to limit the deviation from the precise imitation. When using this update rule, we have observed similar result to those obtained when applying the accurate strategy imitation at large system sizes.

Our simulations were carried out by using $100 \times 100$ system size, but the results remain valid also if we use larger lattices. We implement the model by using synchronous updating, where all the individuals first collect their payoffs through the group interactions and subsequently update their strategies simultaneously. This choice, however, does not limit the validity of our observations because very similar results can be obtained by using asynchronous strategy updating as well. To quantify the cooperative behavior in the population, we compute the cooperation level according to $\rho=L^{-2}\sum_x s_x(\infty)$, where $s_x(\infty)$ denotes the strategy of player $x$ when the system reaches dynamical equilibrium. We also compute the variance of the cooperation level in the equilibrium according to $L^{-2}\sum_x [s_x(\infty)-\rho]^2$. All the results reported in the next section are averages over $100$ independent initial conditions.

\section{Results}

\begin{figure}
\begin{center} \includegraphics[width = 8.4cm]{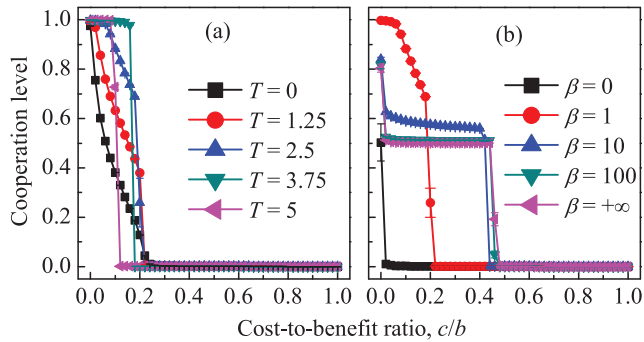}
\caption{\label{cb}(Color online) The average cooperation level and its variance as a function of the cost-to-benefit ratio $c/b$ by using $b=1$. (a) Different values of $T$ (as indicated on the graph) are considered while the steepness is fixed at $\beta=1$. (b) Different values of $\beta$ (as indicated on the graph) are considered while the threshold is fixed at $T=2.5$. The error bars are marked, but they are hardly visible as their size is comparable to that of the symbols.}
\end{center}
\end{figure}

Before presenting the results of the evolutionary process, we note that due to the nonlinearity of the benefit function, higher collective effort from group members will not necessarily result in higher group benefits. This is evident for high $\beta$ values where the collective benefit function $B(S)$ saturates, but may also apply to moderate values of $\beta$. To clarify this point, one can calculate the optimal value of group investments $S$, where the group interest function $P = N b B(S) - Sc$ has a maximum according to
\begin{eqnarray}
S &=& T - \frac{1}{\beta}\ln\frac{1-2y - \sqrt{1-4y}}{2y} \,,\\
\mbox{where}\nonumber\\
y&=&\frac{1}{N\beta}\frac{c}{b} \,.\nonumber
\label{optimal}
\end{eqnarray}
From this it follows that for a real value of $S$, the steepness parameter should be $\beta~>~4c/Nb$. Additional necessary conditions that will result in an optimal group income at $S<N$, however, will depend on the parameters $\beta$, $T$, $S$ and $c/b$ in a non-trivial way.

Turning to the simulation results, we begin by showing in Fig.~\ref{cb}(a) the stationary cooperation level in dependence on the cost-to-benefit ratio $c/b$ for five different values of the threshold $T$ at a fixed steepness $\beta=1$. As expected, the average willingness to contribute to the common pool decreases with increasing $c/b$ for various values of $T$. When the cost-to-benefit ratio is not large, however, maximal investments from players can be achieved at an intermediate value of the threshold. Figure~\ref{cb}(b) shows the cooperation level in dependence on the cost-to-benefit ratio $c/b$ for the fixed threshold $T=2.5$ and five different values of the steepness parameter. It can be observed that the collective investment decreases with increasing $c/b$ for various values of $\beta$. Qualitatively similar as in panel (a), players investment the most at an intermediate value of the steepness when $c/b$ is not large. With increasing $c/b$, the value of $\beta$ that results in the maximum increases. We have also investigated how the variance of the collective investment in the stationary state varies (not shown), finding that it approaches zero for different parameter settings. This implies that the system can fixate into a uniform state where every player contributes to the common with the same rate, \textit{i.e.} where every player adopts the same strategy from the unit interval. As we will elaborate in the continuation of this section, however, the reported fixation may depend on the parameters that characterize the benefit function.

\begin{figure}
\begin{center} \includegraphics[width = 8.4cm]{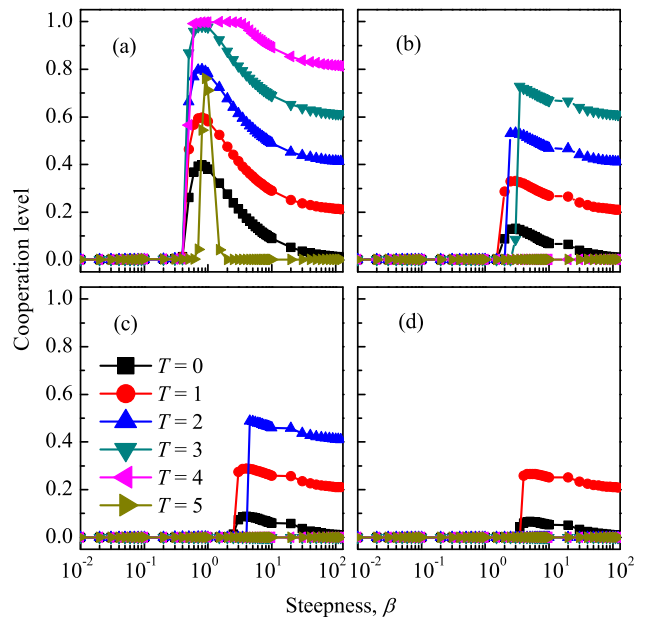}
\caption{\label{cross_s}(Color online) Cooperation level as a function of the steepness parameter $\beta$ for different cost-to-benefit ratios: (a) $c/b=0.1$, (b) $c/b=0.3$, (c) $c/b=0.45$, and (d) $c/b=0.6$. The applied threshold values are indicated in panel (c).}
\end{center}
\end{figure}

In order to explore the impact of the sigmoid benefit function more precisely, we show the cooperation level in dependence on $\beta$ at different values of $T$ for four representative cost-to-benefit ratios in Fig.~\ref{cross_s}. These plots clearly show that there always exists an intermediate value of $\beta$ warranting the best conditions for the selection of the strategy with the highest level of collective contributions within the constraints imposed by any given $c/b$ ratio.

\begin{figure}
\begin{center} \includegraphics[width = 8.4cm]{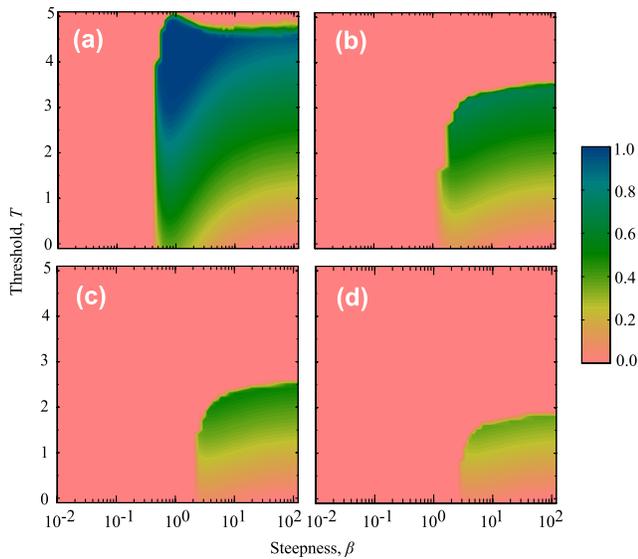}
\caption{\label{plane}(Color online) Contour encoded cooperation level in dependence on the steepness parameter $\beta$ and the threshold $T$ for different cost-to-benefit ratios: (a) $c/b=0.1$, (b) $c/b=0.3$, (c) $c/b=0.45$, and (d) $c/b=0.6$.}
\end{center}
\end{figure}

To provide a more complete view on how the shape of the benefit function influences the evolution of cooperation, we plot the average of stationary $s_x$ values on the whole $T-\beta$ plane in Fig.~\ref{plane} for four different cost-to-benefit ratios. It can be observed that for different values of $c/b$, there always exist an intermediate value of  $T$ that insures the highest collective investment from the population. Furthermore, if the cost of cooperation is not too high, there also exists an intermediate value of $\beta$ that helps the players to maintain their highest level of contributions to the group. Moreover, with increasing the cost-to-benefit ratio $c/b$, the related region of $\beta$ and $T$ shrinks. As a result, the corresponding value of $T$ is decreasing, while $\beta$ value is increasing. Even if $c/b$ is high, \textit{e.g.} $c/b=0.6$, there exist appropriate values of $\beta$ and $T$ that are able to elicit the highest level of collective efforts. However, in such a scenario the average investment first increases from zero to the maximum value, but then for even higher values of $T$ displays little change with increasing $\beta$. Finally, if much higher $c/b$ ratios are used, the optimal region of $\beta$ and $T$ vanishes, and expectedly, full defection reigns in the whole parameter space.

It can be seen in Fig.~\ref{plane} that there always exist a value of $T$ (except for very small $\beta$ values) that ensures the evolution of the highest collective investments. For small threshold values, the amount of produced public goods is high in each group. Thus, everyone can look forward to a high amount of the benefit, and individuals withholding contribution can have a higher payoff than those that do contribute. On the other hand, for high threshold values, the amount of produced public good is small in each group. Thus, everyone obtains only a small amount of the benefit, but individuals withholding contribution can still have higher payoffs than those that do contribute something. Either way, the level of cooperation cannot be high in these two situations. Conversely, for intermediate threshold values, the collective contribution level in some groups can be larger than the threshold, and the amount of produced public goods is hence higher. In this case, individuals with a higher contribution level have the opportunity of collecting higher payoffs than their neighbors with a lower contribution level. Consequently, the former can survive and prevail. In agreement with this insight, there exists an intermediate threshold value that warrants the highest investment from group members. At high $\beta$ values, when the benefit function is practically step-like, the value of $T$ is in strong correlation with the $s_x$ strategy that ultimately prevails. The mechanism that explains this fact, however, will be discussed later in this section.

\begin{figure}
\begin{center} \includegraphics[width = 8.4cm]{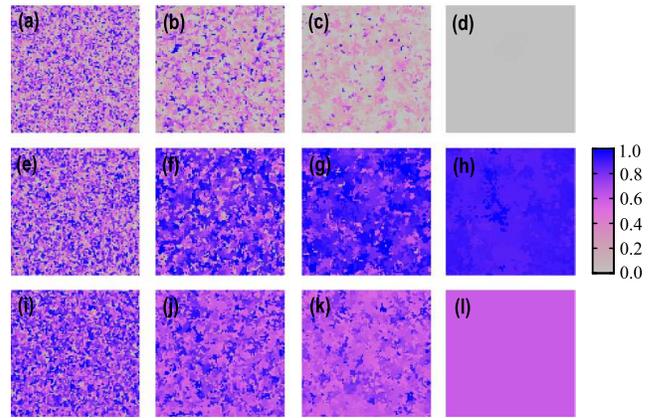}
\caption{\label{3b}(Color online) Evolution (from left to right) of the distribution of individual contributions (strategies) on a square lattice for different values of the steepness parameter: (a,b,c,d) $\beta=0.1$, (e,f,g,h) $\beta=1$, and (i,j,k,l) $\beta=10$. Individuals withholding the entire contribution (pure defectors) are marked gray, while pure cooperators are marked blue. Individual players adopting an intermediate strategy are pink. For further details we refer to the color bar on the right of the figure. Other parameter values are: $c/b=0.1$, $T=2.5$. An identical random initial state was used for three values of $\beta$.}
\end{center}
\end{figure}

\begin{figure*}
\begin{center} \includegraphics[width = 16.0cm]{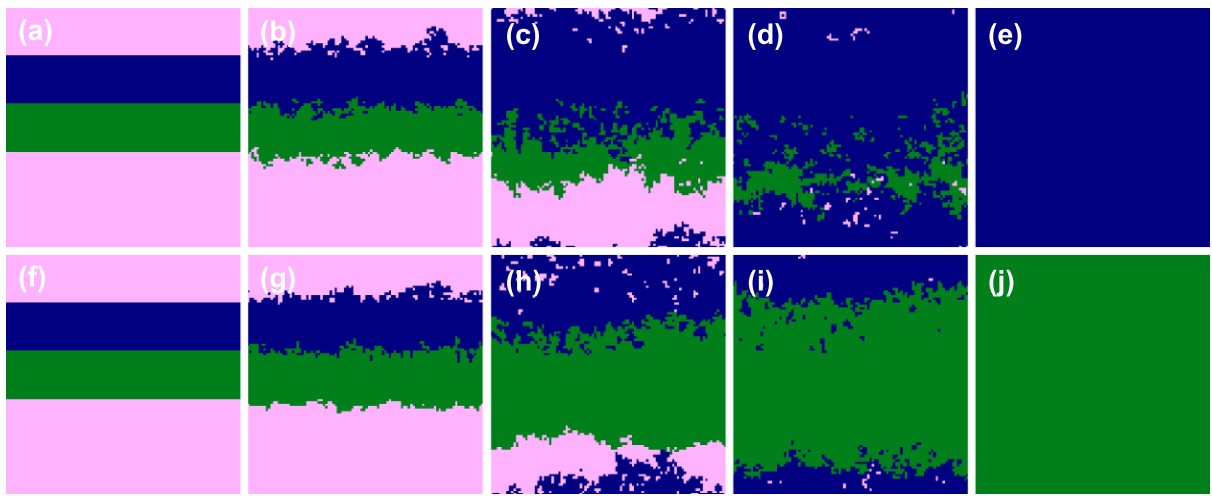}
\caption{\label{front}(Color online) Comparative snapshots depicting the front propagation (from left to right) for the intermediate $\beta=1$ (a,b,c,d,e) and the large $\beta=10$ (f,g,h,i,j) value of the steepness parameter at $T=2.5$. In both rows the initial state contains players having $s_x=0.1$ (pink), $s_x=0.55$ (green), and  $s_x=1$ (dark blue). It can be observed that in the top row both the strategy $s_x=0.55$ and the strategy $s_x=1$ will successfully invade the territory occupied by $s_x=0.1$. Simultaneously, the interface between $s_x=0.55$ and $s_x=1$ is heavily fluctuating and rugged. Conversely, in the bottom row $s_x=0.55$ will dominate not just $s_x=0.1$ but also the $s_x=1$ strategy. Because of this unambiguous superiority, the interface separating $s_x=0.55$ and $s_x=1$ remains comparably smooth and stable. The propagation of this border, however, is slower than that of the one separating $s_x=0.1$ and $s_x=0.55$.}
\end{center}
\end{figure*}

As we have demonstrated, different shapes of the benefit function influence the final output significantly differently. To understand the origins of this better, we plot the time evolution of the spatial distribution of strategies for three representative values of $\beta$ in Fig.~\ref{3b}. In the top row, where a small $\beta$ is applied, players that contribute a lot die out fast. Afterwards, the players that are characterized by a small $s_x$ will also go extinct, and finally full defection will prevail. In this case there are no real consequences related to how large contributions players invest into the common pool because the benefit is virtually independent from it. Given the lack of a substantial advantage, players who have $s_x > 0$ must bear a substantial cost relative to a negligible benefit. Consequently, if the steepness parameter is low defectors will always prevail, regardless of the fact that players are organized in a structured population. At an intermediate steepness [middle row, panels (e) to (h)], full defectors die out first because network reciprocity work in this case: if players cooperate and invest more to the common pool then they can also harvest more, which ultimately results in a competitive individual payoff comparing to the freeriders. In the stationary state, shown in panel (h), different strategies can coexist. Here a delicate balance of investment and cost results in that strategies whose investments are high enough can survive longer. At large $\beta$ [bottom row, panels (i) to (l)], where the prize of mutual investment emerges suddenly, players who contribute less will go extinct soon because they can collect nothing due to the shape of the benefit function. Only those whose investment is large enough to reach the public goods will survive for an intermediate period, but eventually the system fixates into a uniform state where only one strategy remains. Typically this cooperation is approximately consistent with the $S$ value that can be obtained from Eq.~4 and yields the highest collective benefit.

To illustrate the different mechanisms that shape the final output at intermediate and large steepness values, we compare the propagation of fronts separating different strategies (contributing differently to the common pool). For this reason, we have used a prepared initial state containing only three different strategies. In particular, players contribute either $0.1$, $0.55$ or $1$, as described in the caption of Fig.~\ref{front}. In the top row, both $s_x=0.55$ and $1$ will invade against the $s_x=0.1$ strategy because they can both utilize the increasing benefit function. The dominance between the strategies of higher contribution is not so obvious because the higher contribution involves also a higher cost. Consequently, the interface separating these strategies is not smooth as the time evolves, but rather it fluctuates intensively. Albeit, the final state is uniform for this case but the coexistence of strategies is more likely for intermediate values of $\beta$, as we have argued earlier.

In the bottom row, where the benefit function is practically step-like, the relation between the three strategies is more clear. Here the strategy $s_x=0.1$ bares only the cost but experiences no benefits, hence all the other strategies who fulfill $S > T$ can invade it. Accordingly, the stripe populated by $s_x=0.1$ shrinks fast. Although both $s_x=0.55$ and $1$ fulfill the criteria to reap the benefits of a collective investment, the latter players have to bare a larger cost, and consequently $s_x=0.55$ dominate this duel too. The explained superiority between the competing strategies can be observed because the separating fronts remain well determined (the slight fluctuation is the consequence of uncertainty by strategy adoptions). The dominance between $s_x=0.55$ and $1$ is weaker than the dominance between $0.55$ and $0.1$ because the former two strategies differ only in having different costs. This relation can also be observed in Figs.~\ref{front}(g) and (h), where the speed of propagation separating the two borders is different (it is faster between $0.1$ and $0.55$). Finally $0.55$ prevails. The fixation for large $\beta$ values is more likely and subject to the following general scenario: the strategy having the smallest $s_x$ that still fulfills the condition $N S_x~>~T$ to gain the benefit can invade the whole population.

\begin{figure}
\begin{center} \includegraphics[width = 8.4cm]{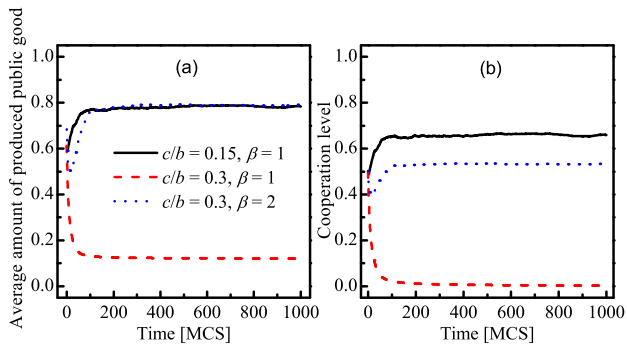}
\caption{\label{time}(Color online) Time evolution of the average amount of produced public goods (a) and the average cooperation level (b) for three different combinations of the cost-to-benefit ratios and the steepness parameter: $c/b=0.15$ and $\beta=1$ (solid black line); $c/b=0.3$ and $\beta=1$ (dashed red line); $c/b=0.3$ and $\beta=2$ (dotted blue line). In all three cases $T=2$.}
\end{center}
\end{figure}

In combination with the above investigations, in the following we explain why the intermediate value of $T$ that warrants the highest collective efforts decreases whereas the related value of $\beta$ increases with increasing $c/b$. In fact, when the cost-to-benefit ratio $c/b$ is increased, the advantage of aggregated individuals with a high level of contribution in collecting payoffs is weakened at large threshold values \cite{szolnoki_pre10}. On the contrary, at lower threshold values the amount of produced public goods can be higher so that individuals who make some contribution to the common pool may have a higher return than those who contribute nothing. Consequently, the cooperation level can be higher in this situation. Furthermore, in Fig.~\ref{time}(a) we observe that, for small values of $c/b$ and $\beta$, the average amount of produced public goods in the population is high, which can provide a high benefit for all the involved individuals. Hence, they can survive and the cooperation level in the population is not low. However, when only the value of $c/b$ is increased, the average amount of produced public good in the population dramatically decreases at the beginning of the evolutionary process. Correspondingly, those individuals that do contribute something to the common pool cannot have a higher payoff than the ones who contribute noting. Eventually, the cooperation level reaches zero [Fig.~\ref{time}(b)]. If the value of $\beta$ is also increased, the average amount of produced public goods can recover to a higher level. Although the positive effect induced by an increased value of $\beta$ is still restricted by the higher $c/b$ ratio, the final stationary cooperation level can still reach a relatively higher level.

\begin{figure}
\begin{center} \includegraphics[width = 8.4cm]{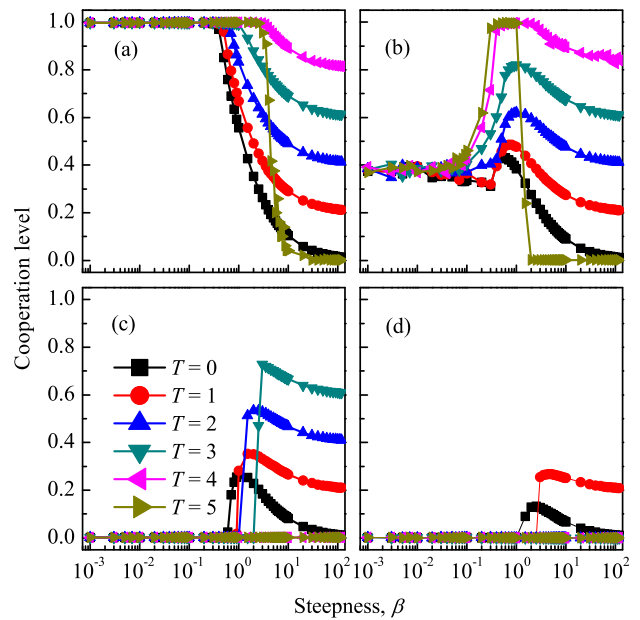}
\caption{\label{cross_c}(Color online) Cooperation level as a function of the steepness parameter $\beta$ for different cost-to-benefit ratios: (a) $c/b=0.1$, (b) $c/b=0.2$, (c) $c/b=0.3$, and (d) $c/b=0.6$. The applied threshold values are indicated in panel (c). Importantly, unlike in Fig.~\ref{cross_s}, here the more complex benefit function is applied, as defined by Eq.~\ref{compl}.}
\end{center}
\end{figure}

Finally, to explore the robustness of our findings, we consider a modified yet more complex form of the benefit function following previous work \cite{archetti_ev11}:
\begin{equation}
B(S_i)=\frac{W(S_i)-1}{W(N)-1},
\label{compl}
\end{equation}
where
\begin{equation}
W(z)=\frac{1+exp(\beta T)}{1+exp[-\beta(z-T)]}.
\end{equation}
It is worth emphasizing that this modified form of the benefit function is still sigmoid. When $\beta \rightarrow 0$, the produced public good is a traditionally linear function of individual contributions; when $\beta\rightarrow +\infty$, the produced public good is a step-like function of the individual contributions. In other words, this modified sigmoid function is strictly constrained between the linear and the step-like shape. Figure~\ref{cross_c} presents the cooperation level in dependence on the steepness $\beta$ for different values of the threshold $T$ for four different situations: (a) $c/b=0.1$, (b) $c/b=0.2$, (c) $c/b=0.3$, and (d) $c/b=0.6$. We see that even under the action of an alterative benefit function, there still exists an intermediate value of $T$ that warrants the highest investments from players. When the $c/b$ ratio is increased, the optimal value of $T$ is reduced and the cooperation level also decreases. Whereas for small values of $c/b$, \textit{e.g.} $c/b=0.1$, the cooperation level monotonously decreases with increasing the steepness $\beta$ for different values of $T$. However, when $c/b$ becomes larger, \textit{e.g.} $c/b=0.2$, there also exists an intermediate value of $\beta$ warranting the best promotion of cooperation for different values of $T$. As $c/b$ continues to increase, the non-monotonous dependence of the cooperation level on $\beta$ only occurs at smaller values of the threshold, and the largest cooperation level is correspondingly reduced. For still larger $c/b$ ratios, full defection is reached irrespective of the threshold and the steepness parameters (not shown here), likewise as reported above for the originally considered benefit function.

\section{Discussion}
Summarizing, we have studied the evolution of cooperation in the spatial continuous public goods game subject to different sigmoid benefit functions. We have shown that there exists an intermediate threshold value as well as an intermediate steepness, at which the collective contributions to the common pool are the largest. Upon increasing the cost-to-benefit ratio, we have found that the threshold value related to this maximum decreases, while the corresponding measure of steepness increases. Simultaneously, the parameter region where public cooperation can prevail was found shrinking, and ultimately vanishing completely at a critical cost-to-benefit ratio. When employing a more complex variant of the sigmoid functional form in order to describe the governing benefit functions, we have discovered that our results remain robust, \textit{i.e.} there always exist intermediate values of the threshold and the steepness at which investments to the common pool are the largest. As by the usage of the simpler benefit functions, in case of more complex variants the related specific values of both parameters were also found to depend significantly on the cost-to-benefit ratio. Taken together, our results elucidate the impact of generalized benefit functions on the evolution of cooperation in the spatial public goods game, which appear to always enable the tragedy of the commons to be avoided.

This work continues along the lines of previous investigations considering different benefit functions, yet it does so on structured rather than well-mixed populations. Unlike in well-mixed populations \cite{archetti_ev11}, however, here we find that the outcome of the public goods game depends significantly on the steepness parameter, and that thus the step-like benefit function is not necessarily a good approximation for an arbitrary nonlinear benefit function. In fact, due to spatial reciprocity the sigmoid benefit function can significantly broaden the domain where cooperative behavior can survive even by relatively unfavorable cost-to-benefit ratios. The presented results thus promote our understanding of the effects of nonlinear benefit functions on the evolution of public cooperation, especially if spatial reciprocity is a contributing factor.

In comparison to \cite{szolnoki_pre10}, where the impact of the critical mass, \textit{i.e.} the threshold number of cooperators that is required for harvesting the benefits of the collective effort, was considered in the spatial public goods game with two discrete strategies, we adopt here a more generalized approach, where instead of just the threshold also the steepness is considered as a free parameter. Moreover, instead of the two discrete strategies, we consider the full continuous array of strategies allowing for the delicate variability of contributions to the common pool. In this broader framework, we confirm that moderate threshold values can warrant the highest mutual contributions from all the group members, albeit depending on the value of the cost-to-benefit ratio and the steepness. More importantly, we find that there exists an intermediate steepness of the benefit function, which can further amplify the positive effects of an appropriately adjusted threshold. Our results thus convey the possibility of a double enhancement of collective contributions, thus highlighting the important role of nonlinear benefit functions for the evolution of prosocial behavior on structured populations.

\begin{acknowledgments}
Financial support from the Hungarian National Research Fund (Grant No. K-73449), the Slovenian Research Agency (ARRS) (Grant No. J1-4055), the 973 Program (Grant No. 2012CB821203) and the National Natural Science Foundation of China (NSFC) (Grants Nos. 61020106005 and 10972002) is gratefully acknowledged.
\end{acknowledgments}

\end{document}